\documentclass[10pt,a4paper,prl,twocolumn]{revtex4}
\usepackage{amsmath}
\usepackage{amsfonts}
\usepackage{amssymb}
\usepackage{graphicx}
\begin{document}
\title{Dynamically induced effective interaction in periodically driven granular mixtures}

\author{Massimo Pica Ciamarra}\email[]{picaciamarra@na.infn.it}
\author{Antonio Coniglio}
\author{Mario Nicodemi}

\affiliation{Dipartimento di Scienze Fisiche, Universit\'a di
Napoli `Federico II', CNR-Coherentia, INFN, 80126 Napoli,
Italia.}
\homepage{http://smcs.na.infn.it}

\date{\today}

\begin{abstract}
We discuss the microscopic origin of dynamical instabilities and segregation
patterns discovered in granular mixtures under oscillating horizontal shear, by
investigating, via molecular dynamics simulations, the effective
interaction between like-particles. This turns out to be attractive
at short distances and strongly anisotropic, with a longer range
repulsive shoulder along the direction of oscillation. This features
explain the system rich phenomenology, including segregation and
stripe pattern formation. Finally, we show that a modified
Cahn-Hilliard equation, taking into account the characteristics
of the effective interaction, is capable of describing the dynamics
of the mixture.
\end{abstract}
\maketitle

Dynamical instabilities and pattern formation, so important in Fluid 
Mechanics, have just begun to be discovered in granular materials~\cite{shinbrot,noi_prl},
which are collections of macroscopic particles interacting via dissipative forces where thermal effects 
are negligible.
Understanding their origin in these systems, which is crucial for scientific reasons and substantial 
industrial applications, is today a challenge as granular systems cannot be
simply described by usual Statistical or Fluid Mechanics
\cite{shinbrotrev,richard,segregation_review,noi_prl,noi_press,prox}.
In this Letter we make a step further in this direction by
showing that the study of peridocially driven binary granular
mixture can be reduced to that of a thermal monodisperse undriven system
of particles interacting via an effective potential. 
This result can be considered as an extension, to the non-thermal and driven case, of 
the `depletion potential' approach introduced by Asakura and Oosawa~\cite{asakura}
to map an undriven thermal binary mixture in a monodisperse system.
However, while the depletion potential has a purely entropic origin as 
related to the size difference of the two components, the
effective interaction we introduce here has a purely dynamical origin, 
as it results from the different response of the 
mixture components to the oscillating drive. Our approach could be also
of value in the study of driven thermal binary mixutres..

We apply our ideas to a granular mixture subject to horizontal oscillations, 
a system previously investigated both experimentally and 
numerically~\cite{mullin,noi_prl,yeomans,king}, whose complex and not
well understood phenomenology encompasses both instabilities and segregation.
First we determine, via Molecular Dynamics (MD) simulations, the effective interaction
between like-particles of this mixutre when submitted to horizontal oscillations. 
Then we show that simulations of just one species of particles interacting via this effective interaction,
in absence of horizontal oscillations, reproduces the phenomenology of the oscillated granular mixture. 
Finally, guided by the main properties of the effective interaction between like particles, we introduce
a phenomenological Cahn-Hilliard equation which again reproduces the observed phenomenology,
but also allows for analytical predictions. 

{\it Model --} We have investigated via MD
simulations a two-dimensional model~\cite{noi_prl} of the experiment
of Ref.~\cite{mullin}, where a monolayer of granular mixture is
placed on a horizontally oscillating tray of size $160D\times 40D$
($D=1$cm). In our model, contacting particles interact via a spring
dashpot repulsive force with constant coefficient of restitution $e
= 0.8$, and the interaction between a grain and the oscillating tray
is given by ${\bf f}_{\rm tray} = -\mu ({\bf v}-{\bf v}_{\rm
tray})$, where ${\bf v}_{\rm tray}(t) = 2\pi A \nu \sin(\nu t) {\bf
x}$ is the velocity of the tray and ${\bf v}$ the velocity of the
disk, plus a white noise force ${\bf \xi}(t)$ with $\langle {\bf
\xi}(t) {\bf \xi}(t') \rangle = 2\Gamma \delta(t-t')$ (see
\cite{noi_prl,Silbert2001} for details). 
The two components of our mixture have masses $m_{\rm h} = 1$ g, $m_{\rm l} = 0.03$ g, viscous
coefficient $\mu_{\rm h} = 0.28$ g s$^{-1}$ and $\mu_{\rm l} = 0.34$
g s$^{-1}$,  the white noise has $\Gamma = 0.2$ g$^2$cm$^2$s$^{-3}$,
and the tray oscillates with amplitude $A = 1.2$ cm and frequency
$\nu = 12$ Hz. 
The diameters of the two species considered here are
equal, $D_h = D_l=D=1$ cm, to remark that no ``entropic''  depletion
forces are present.

{\it Effective interaction --} 
In order to determine the effective interaction we have performed
simulations where two heavy disks are placed at fixed relative
positions ${\bf r}_{12}= {\bf r}_1-{\bf r}_2 =(x,y)$ in a system of lighter
disks covering an area fraction $\phi \simeq 0.63$ (we have sampled the range $0 <x,y < 6D$). 
When the tray oscillates
horizontally the two heavy disks are moved as a single object of
mass $M = 2m_{\rm h}$ subject to a force ${\bf f} = {\bf f}_1 + {\bf
f}_2 + (2 {\bf f}_{\rm tray} + 2 {\bf f}_{\rm noise})$, where ${\bf
f}_1$ (${\bf f}_2$) is the force acting on particle $1$ ($2$) due to
collisions with lighter disks. The two disks can translate
horizontally and vertically, but ${\bf r}_{12}$ remains fixed. The
effective force that particle $1$ exerts on particle $2$, averaged
over one period of oscillation $T$, is given by
\begin{equation}
{\bf f}^{\rm eff} = (f^{\rm eff}_x,f^{\rm eff}_y) = \frac{1}{2T}\int_0^T[{\bf f}_1(t) - {\bf f}_2(t)]dt.
\end{equation}
In the study of different systems, as for instance for two spheres in a oscillating fluid flow, 
an analytical expression for the effective interaction force can be in principle found.

\begin{figure}[t!!!]
\begin{center}
\includegraphics[scale = 0.3]{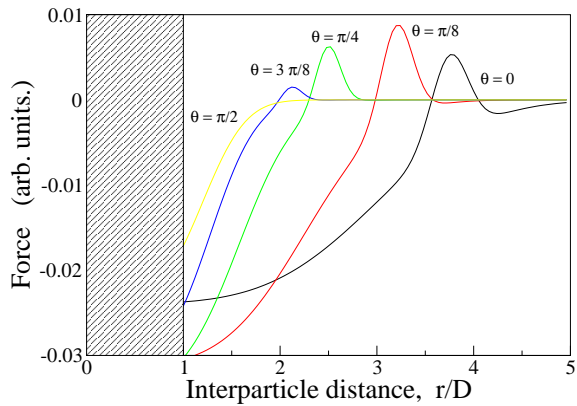}
\end{center}
\caption{\label{fig-force} (color on line)
Radial component of the effective force along directions forming an angle $\theta$ with the $x$ axis.
For $r/D < 1$ the force is strongly (out of scale) repulsive due to steric constraints.
}
\end{figure}

Our numerical results show that this force is attractive at short
distances, and strongly asymmetrical in the $xy$-plane: it has a
repulsive shoulder at long distances, which vanishes along the $y$
axis. This is shown in Fig.~\ref{fig-force}, where we plot the
radial component of the effective interaction force along
directions forming an angle $\theta$ with the $x$ axis.
The effective force is weakly dependent on $\Gamma$ 
(in the investigated range $0 < \Gamma < 10$~g$^2$cm$^2$s$^{-3}$),
and its amplitude increases with the area fraction of the smaller species.
Variations of the amplitude and frequency of oscillation do not
change the qualitative features of the effective force (asymmetry,
attraction at short distances and repulsive shoulder), but they
change the range of attraction and the position of the repulsive
shoulder. Rigourously, the effective force ${\bf f}^{\rm eff}$
cannot be derived by an effective scalar potential, as the curl of
${\bf f}^{\rm eff}$ ($\nabla \times {\bf f}^{\rm eff} =
\partial_y f^{\rm eff}_x - \partial_x f^{\rm eff}_y$) 
varies in space (the presence of a solenoid component in the effective interaction has been
also reported in~\cite{dzubiella} for a system of two disks in a constant fluid flow).
However, as the solenoid component of ${\bf f}^{\rm eff}$ appears to be negligibe with respect to
its irrotational component~\footnote{According to Helmholtz's theoreom any vector field ${\bf V}$ admits a unique decomposition in a irrotational and solenoidal component, ${\bf V} = {\bf V}_I + {\bf V}_S$. The solenoidal component
of ${\bf f^{\rm eff}}$ appears to be negligible with respect to its irrotatioanl component, ${\bf f}_S(r) \ll {\bf f}_I(r)$.}, an effective interaction scalar potential can be introduced within a good approximation.

\begin{figure}[t!!!]
\begin{center}
\includegraphics[scale = 1]{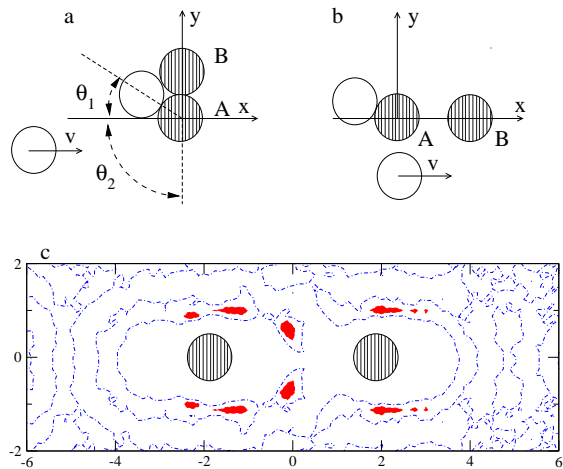}
\end{center}
\caption{\label{fig-density} (color on line) 
The effective interaction between two heavy disks is determined
via screening effects by their relative position, as discussed in the text
an exemplified in Panels a,b. Panel c is a contour level plot of the density field 
of light disks in an oscillating system. It shows that when the distance between the
two heavy disks along the direction of oscillation increases light disks are caged between 
them giving rise to an effective repulsion (see text).
Panel c has been obtained by recording the position of all particles after $1/4$ of 
oscillation in the frame of reference centered in the 
mid-points between the two heavy disks (kept at the fixed distance of $3.75~D$), 
and averaging over $100$ configurations. 
The regularity of the isodensity lines is due to steric constrains. 
Filled area represent regions of density above the average.
}
\end{figure}

The effective force we derive (see Fig.\ref{fig-force}) appears to be the key ingredient 
in understanding the phenomenology of our system. 
It follows from the different response of the two species of particles 
to the oscillating drive (the tray), which are forced to oscillate with different amplitues and phases
due to their differences in mass and friction coefficient.
A qualitatively understanding of the microscopic origin of the effective force
can be obtained by considering the simpler case where two disks of a given species (striped disks
in Fig.~\ref{fig-density}a,b) are immersed in stream of disks of a different species 
flowing along the $x$ axis with velocity $v > 0$. 
Fig.~\ref{fig-density}a illustrates that, as $\theta_1 < \theta_2$, the majority of the collisions experienced by grain A push it closer to grain B, explaining the
attraction between particles A and B along $y$.
Fig.~\ref{fig-density}b shows that, for small distances, particle A screens particle B, 
inducing an effective attraction between them along $x$. For larger distances,
and in the presence of the oscillating drive, light disks tend to be caged
between the two heavy ones, and their density may become higher
than average, as shown in Fig.~\ref{fig-density}c. 
In this condition the competition between the collisions experienced by the
heavy disks from the caged light disks, which push them apart,
and those experience from the surrounding disks, which push them toghether,
is won by the caged disks and results in an effective repulsive force.

{\it Segregation and instability --} The effective force, ${\bf f}^{\rm eff}$, 
was derived for a system of just two heavy grains in a
``bath'' of lighter ones. For a system with many heavy grains 
this is, in general, expected to be just 
an approximation. We checked, however, that ${\bf f}^{\rm eff}$
captures the basic physical mechanism responsible for segregation 
via stripes formation in the investigated system. To this aim 
we run simulations of a monodisperse system of
heavy disks only, where disks interact pairwise via the effective
force previously determined. As in the original mixture the disks
are also subject to a viscous force (${\bf f}_{\rm tray} = -
\mu_{\rm h}{\bf v}$) and to a white noise force ${\bf \xi}$. They
are placed on a \textit{fixed} tray of size $160D\times 40D$ 
and cover an area fraction $\phi \simeq 0.5$.

\begin{figure*}[t!!]
\begin{center}
\includegraphics[scale = 0.28, angle = -90]{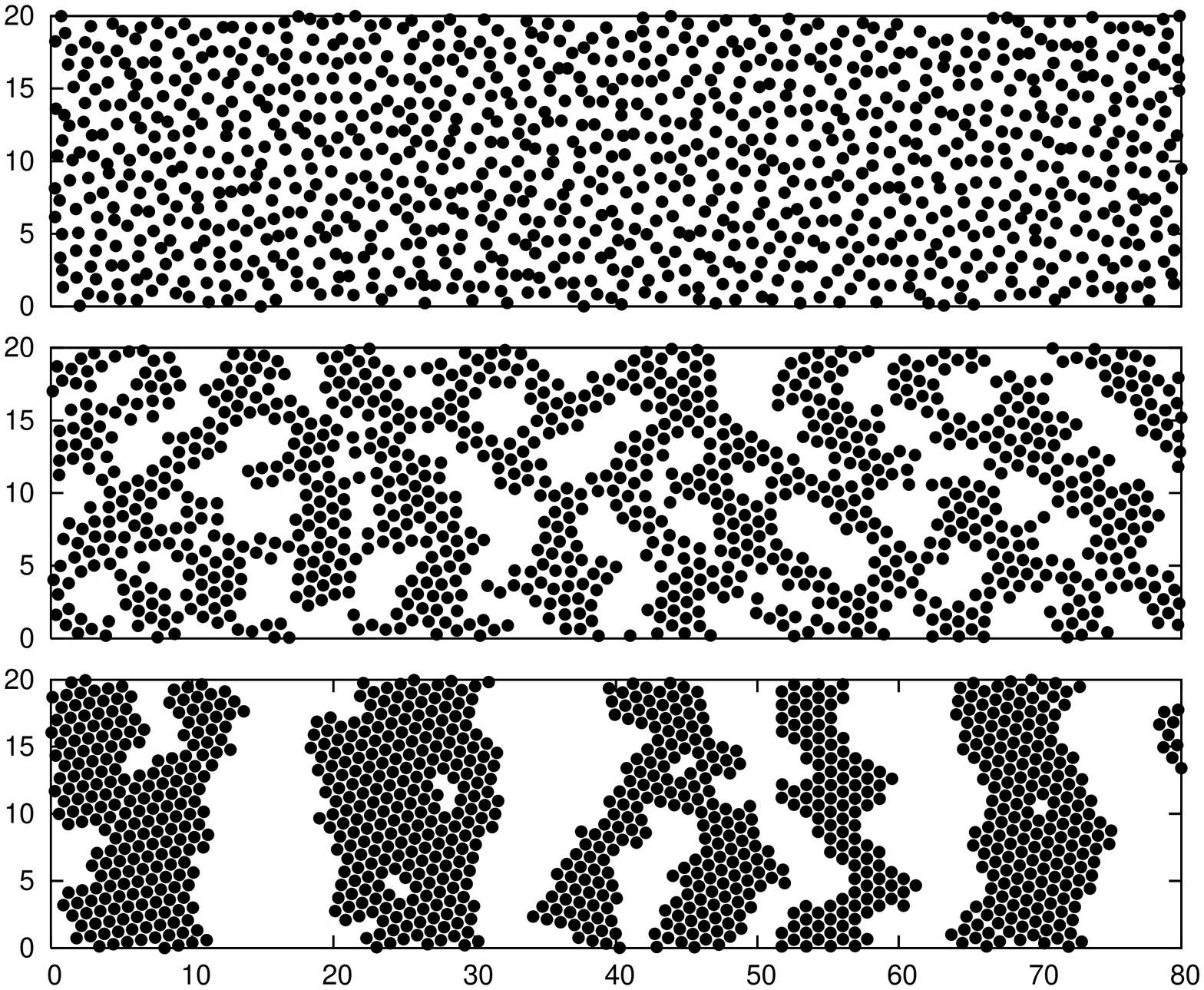}~
\includegraphics[scale = 0.28, angle = -90]{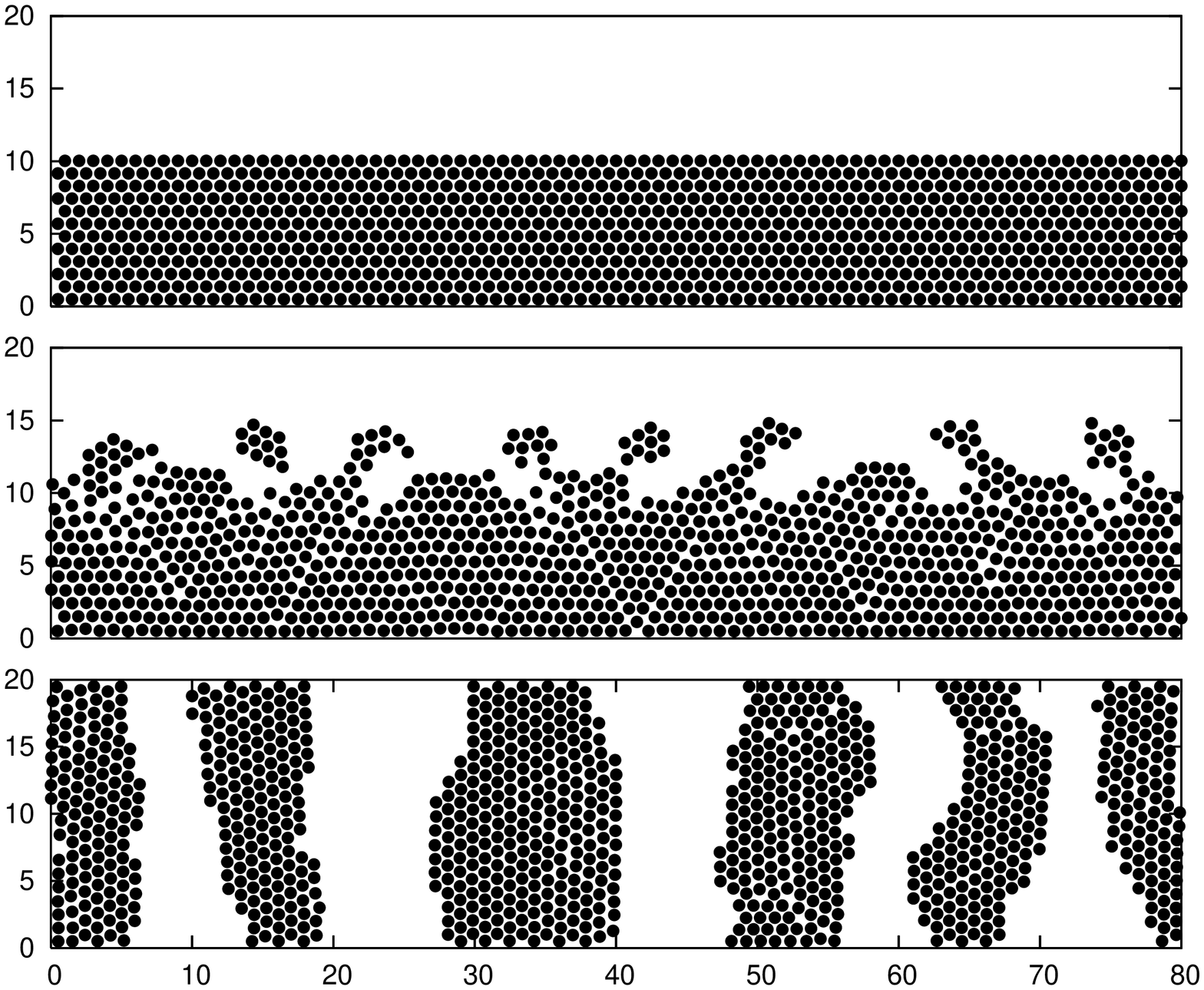}
\end{center}
\caption{\label{fig-evolution} 
Evolution of a system of disks interacting via the effective force shown in
Fig.~\ref{fig-force}. Lenghts are expressed in units of particles diameters.
Left column (periodic boundary conditions in both direction): from top to bottom, 
evolution from an intially disorder state at $t = 0,4$ and $100$s.
Right column (periodic boundary condition along $x$, elastic along
$y$): evolution from an initially segregated state at $t=0,8$ and $300$s. 
Each single component of a real granular mixtures subject to
horizontal oscillations
evolves in a similar way~\cite{noi_prl}.}
\end{figure*}
Fig.~\ref{fig-evolution} shows that this system evolves, from an
initially disordered configuration, via the formation of
interconnected clusters that at long times break into a pattern of
stripes parallel to the $y$ direction. This is precisely the
behavior exhibited by each single component of the original
horizontally oscillated binary mixture of disks in simulations (see
Fig.1b of Ref.~\cite{noi_prl}) and experiments~\cite{mullin}. An
insight on the system behavior can be gained via an analogy with
thermal systems with short range attraction and long range
repulsion~\cite{glotzer}. In these systems the short range
attraction, which tends to induce a macroscopic phase separation
with the formation of a single large cluster in the system, is
frustrated by the presence of the long range repulsion 
adversing large clusters. Depending on the relative strength of
attraction and repulsion, the attraction may dominates and the
system will eventually phase separate via the standard coarsening
mechanism, otherwise if the repulsion dominates the coarsening
process eventually stops when a typical domain width is reached,
resulting in the formation of striped patterns~\cite{glotzer}. 
In our granular system, as the long range repulsive component of the effective force
vanishes along the $y$ axis (see Fig.~\ref{fig-force}, $\theta =
\pi/2$), cluster growth along $y$ is not frustrated. Along $x$
the relative strength of the repulsion is small compared to the attraction, and
the coarsening process will therefore proceeds slowly  
as long as the external driving keeps going. The asymmetry
of the interaction force therefore explains the formation and the
orientation of the striped pattern.

At low values of the strength of the white noise force,  $\Gamma$, the
stripes appears to be formed by disks in an ordered (crystal) state (see
Fig.~\ref{fig-evolution}). At a higher value of $\Gamma$ the stripes
appears fluid-like, while at stiller higher values the system does
not segregate. The same behavior is observed in a real mixture
of disks subject to horizontal oscillation both when $\Gamma$ is
increased, and when the area fraction of one of the two components
is decreased~\cite{mullin,noi_prl}.

The effective interaction explains as well the occurrence of the
dynamical instability generating the above segregated pattern~\cite{noi_prl}.
The instability is best visualized when
the initial state of the system is not disordered, but disks interacting with 
the effective force are placed on the tray in a stripe parallel to the $x$ direction (see Fig.~\ref{fig-evolution}).
In this condition the initially flat free surface 
develops a sine-like modulation which grows until it
breaks giving rise to a pattern of alternating stripes perpendicular to $x$ direction. 
The same phenomenology is observed when the two components of the granular mixture
are placed on the oscillating tray in two stripes parallel to the $x$ direction (see, for instance,
Fig.1a of Ref.~\cite{noi_prl}).

\begin{figure*}[t!!!]
\begin{center}
\includegraphics[angle = -90, scale=0.55]{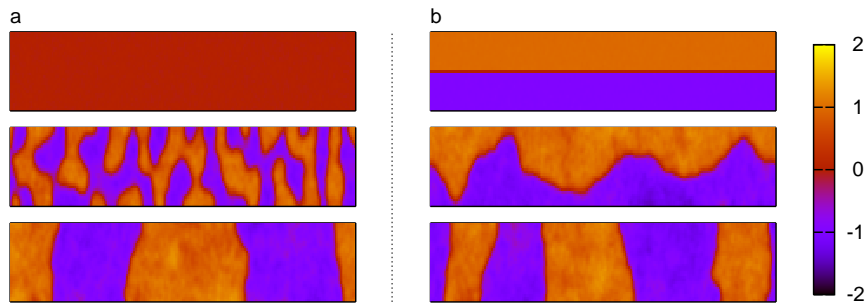}
\end{center}
\caption{\label{fig-ch} (color on line) Solution of the modified
Cahn-Hilliard equation (Eq.~\ref{eq-ch}),
with $K_y/2 = K_x = 1$ in a system of size $200 \times 32$ with
periodic boundary conditions along $x$ and no flux boundary conditions
along $y$. (a): evolution from an homogeneous initial condition shown
at times $t=0,100,2500$; (b) evolution from a segregated initial
condition shown at times $t=0,750,2500$. The color scale on the right
is a measure of the density difference field.}
\end{figure*}

{\it Cahn-Hilliard approach --}
As the Cahn-Hilliard equation captures the general features of 
spinodal decomposition of thermal binary mixtures (see~\cite{goldenfeld} for a review),
regardless of the details of the interaction potential between the two
components, we expect that a phenomenological Cahn-Hilliard equation
may capture the properties of our granular system when coarsening is observed.
Therefore we have investigated a Cahn-Hilliard equation for the density difference
$c(\vec r) = \rho_1(\vec r) - \rho_2(\vec r)$ of the two components, which takes into account the
anisotropy found in the effective force field
\begin{equation}
\frac{\partial c(\vec r)}{\partial t} = M\nabla^2\left[
\frac{\partial f}{\partial c}-(K_x\partial_x^2+K_y\partial_y^2)c\right]+\eta.
\label{eq-ch}
\end{equation}
Here $M$ is a mobility and $\eta$ a Gaussian random noise~\cite{nota1}, 
$f(c)$ the usual double-well potential leading to phase separation, 
whereas the term $K_x\partial_x^2+K_y\partial_y^2$ accounts for the
``free energy'' cost associated to concentration gradients. 
In the present case, as the effective interaction is not spherically
invariant, the cost of an interface depends on its orientation.
Schematically we take this into account by assuming $K_y > K_x$ (in
case of radial isotropy $K_x = K_y$) since interfaces 
(i.e., concentration gradients) along $y$
``cost more'' than interfaces along $x$.
We show in Fig.~\ref{fig-ch}a, which consider the case in which the
initial state is homogeneous, and in Fig.~\ref{fig-ch}b, which
reproduces the instability of an initially flat interface between the
two components, that this phenomenological equation gives a good
description of the dynamics. A quantitative agrement is also found as 
the Cahn-Hilliard equation predicts a coarsening
exponent $\nu = 1/4$ during the first coarsening regime~\cite{zannetti},
which is numerically equal to the one observed in both experiments and
simulations~\cite{mullin,prox}.
According to the Cahn-Hilliard 
equation we expect, however, a crossover~\cite{zannetti} to
$\nu = 1/3$ at much longer times, not yet
observed in experiments and simulations. 

{\it Conclusions --}
In this Letter we have generalized the effective interaction approach, widely used in the study
of colloidal systems, to out-of equilibirum periodically driven mixtures, and specifically to
the case of a granular mixture subject to horizontal oscillations. In this context the effective
interaction approach is particularly useful, as it reduces the study of an out-of equilibrium non-thermal
driven system (dissipative in the case we have explicitely investigated), to that of an ``equilibirum''
 monodisperse system. For a granular mixture subject to horizontal oscillations, the effective interaction force, 
whose foundamental chracteristics is its directional anisotropy, and particularly the presence of a 
repulsive shoulder at long distances which is prominent in the direction of oscillation, allows for a clear understanding of the observed instabilities and segregation processes. Its features can be casted
in a phenomenological Cahn-Hilliard equation which reproduces the observed phenomenology and allows for analytical predictions.
Our findings clarify, thus, the origin of the ``differential drag'' mechanism~\cite{king} 
proposed to describe the observed phenomena and show how
to interpret phenomenological hydrodynamics models~\cite{yeomans} used
to depict the early stages of the stripe dynamics. 

\begin{acknowledgments}
Work supported by EU Network Number MRTN-CT-2003-504712, MIUR-PRIN 2004,
CrdC-AMRA.
\end{acknowledgments}


\begin{thebibliography}{40}
\bibitem{nota1} The Gaussian random noise $\eta$ is obtained as the
divergence of a random vector field in order to keep the order
parameter conserved.
\end{thebibliography}
\end{document}